\documentclass[reprint, nofootinbib, amsmath, amssymb, aps, prl]{revtex4-2}

\usepackage[pdftex]{graphicx, color}

\usepackage{physics}
\usepackage{eucal}
\usepackage{bbm}
\usepackage{ulem}

\begin{document}

\title{
    More fields are different: \\ 
    ---Stochastic view of multi-field inflationary scenario---
}

\author{Tomo Takahashi}%
\email{tomot@cc.saga-u.ac.jp}
\affiliation{%
	Department of Physics, Saga University, 1 Honjomachi, Saga 840-8502, Japan
}%

\author{Koki Tokeshi}%
\email{tokeshi@icrr.u-tokyo.ac.jp}
\affiliation{%
	Institute for Cosmic Ray Research (ICRR), The University of Tokyo, 5-1-5 Kashiwanoha, Kashiwa, Chiba 277-8582, Japan
}%

\date{\today}

\begin{abstract}
\vspace{4pt}\noindent
High-energy physics often motivates multi-field inflationary scenarios where stochastic effects play a crucial role. 
Peculiar to multi-field models, the noise-induced centrifugal force results in a longer duration of inflation depending on the number of fields, even when the stochastic noises themselves are small. 
We show that, in such small-noise regimes, the number of fields generically discriminates whether inflation successfully terminates or lasts forever. 
Our results indicate that inflation with an extremely large number of fields may fail to realise our observable Universe. 
\end{abstract}

\maketitle 

\noindent\textit{\bfseries Introduction.}---Cosmic inflation~\cite{Starobinsky:1980te, Sato:1980yn, Guth:1980zm, Linde:1981mu} is now the leading paradigm that describes the earliest epoch of the Universe. It enables every pair of points in our observable Universe into a single causal region, eliminating the need for fine-tuned initial conditions in the standard Big Bang scenario. Vacuum quantum fluctuations generated during inflation were stretched and crossed out the horizon to behave classically, which were later amplified by gravitational instability to give rise to all the present cosmological structures such as stars and galaxies. 

While various large-scale observations have been confirming the nearly scale-invariant power spectrum predicted by the simplest inflationary model realised by a single scalar field~\cite{Planck:2018vyg, Planck:2018jri}, it seems more natural to have a multi-field scenario especially from the point of view of particle physics models based on supersymmetry, supergravity, or string theory, often characterised by a large number of scalar fields. In addition to the compatibility with high-energy model constructions, a class of those multi-field models is attractive not only because it has a richer structure than single-field models such as generation of non-adiabatic fluctuations as well as non-Gaussianities \textit{e.g.}~\cite{Wands:2007bd, Langlois:2010xc, Byrnes:2010em, Takahashi:2014bxa}, but also it in some cases can realise a sufficient duration of inflation while keeping the inflatons' excursion in a sub-Planckian regime~\cite{Liddle:1998jc, Kanti:1999ie, Dimopoulos:2005ac, Kim:2006ys}. Various possibilities of a wide range of the number of fields, depending on models and requirements, have been discussed so far, see \textit{e.g.}~\cite{Piao:2002vf, Brandenberger:2003zk, Ahmad:2008eu, Battefeld:2008qg}. 

When a large number of fields is prevalent during inflation, however, stochastic effects may play an important role even when the stochastic noise coming from quantum fluctuations is small. Accumulation of scalar fields realise another source of the deterministic, but stochastic noise-induced, force, which is in contrast to single-field models where the inflaton solely descends the potential under the classical drift force except a rare realisation of the stochastic noise. Such stochastic effects can be taken into account by the stochastic description of inflation formulated in~\cite{Starobinsky:1986fx} and later systematically developed to extend its applicability to multi-field models in~\cite{Assadullahi:2016gkk}. 

To study multi-field dynamics in stochastic inflation is however difficult even numerically, since it consists of stochastic and partial differential equations. Meanwhile, there is a class of models of particular interest, such as those with an $O (d)$-symmetric potential with $d$ fields, which has been investigated in~\cite{Assadullahi:2016gkk, Vennin:2016wnk, Tada:2023fvd}. The statistical moments of the number of $e$-folds, under the stochastic effects, are the prime quantities since they are directly related to the observables such as the curvature perturbation and its power spectrum through the non-perturbative relation, nowadays called the stochastic-$\delta \mathcal{N}$ formalism~\cite{Fujita:2013cna, Fujita:2014tja, Vennin:2015hra}. 

The present article is interested in how and how much the number of fields $d$ affects the mean number of $e$-folds elapsed during inflation, subjected to the stochastic effects. We consider a class of $O(d)$-symmetric models to demonstrate that the duration of inflation can be infinite, known as \textit{eternal inflation}~\cite{Vilenkin:1983xq, Linde:1986fc, Linde:1986fd}, depending on $d$ even when the stochastic effects are small. The requirement that inflation terminates to realise our observable Universe puts a theoretical bound on $d$ from above. 
This gives significant implications for the early-Universe scenarios motivated by high-energy physics, since inflation inevitably lasts forever if $d$ is too large unless the initial condition is finely tuned. Such an infinite duration of inflation comes from the fact that not only the classical drift but also the stochastic noise-induced centrifugal force affects the mean number of $e$-folds. The latter is peculiar to multi-field models, due to which the inflaton fields not only descend but also ascend on the potential. The effective potential that takes the centrifugal force into account enables us to generically identify the condition for the number of $e$-folds to be finite or infinite, which depends on $d$ and the initial condition of the fields. Throughout this article, natural units are used and $M_{\rm P}$ denotes the reduced Planck mass. 

\vspace{8pt}\noindent
\textit{\bfseries Stochastic multi-field inflation.}---During inflation, small-scale quantum fluctuations are stretched and cross out the horizon to be classicalised. The large-scale modes are then subjected to a continuous and random inflow from the small-scale fluctuations. This transition from small-to-large scales is elegantly described by the stochastic formalism of inflation~\cite{Starobinsky:1986fx} by decomposing the entire fields into those two modes, 
\begin{align}
	\vb*{\phi} (N, \, \vb*{x}) 
	&= \vb*{\phi}_{-} (N, \, \vb*{x}) + \vb*{\phi}_{+} (N, \, \vb*{x}) 
	\,\, , 
	\label{eq:decomp}
	\\ 
	\vb*{\phi}_{\pm} (N, \, \vb*{x}) 
	&\equiv \int \frac{\dd^{3} k}{(2 \pi)^{3}} \, \Theta \qty[ \pm k \mp k_{\sigma} (N) ] \widetilde{\vb*{\phi}} (N, \, \vb*{k}) e^{i \vb*{k} \cdot \vb*{x}} 
	\notag 
	\,\, ,
\end{align}
where $\vb*{\phi} = (\phi_{1}, \, \dots, \, \phi_{d})^{\mathsf{T}}$ is a set of the $d$ scalar fields and $k_{\sigma} \equiv \sigma a H$ is the time-dependent cutoff scale ($0 < \sigma \ll 1$). The number of $e$-folds $N$ is used as the time variable instead of the cosmic time $t$~\cite{Vennin:2015hra}, defined through the Hubble parameter by $\dd N = H \, \dd t$. The effective equation of motion for the coarse-grained field $\vb*{\phi}_{-}$, denoted $\vb*{\phi}$ hereafter, can then be obtained by integrating out the small-scale modes $\vb*{\phi}_{+}$, 
\begin{equation}
	\dv{\vb*{\phi}}{N} 
	= - \frac{ \nabla V (\vb*{\phi}) }{3 H^{2} (\vb*{\phi})} + \frac{H (\vb*{\phi})}{2 \pi} \vb*{\xi} (N) 
	\,\, , 
	\label{eq:lan}
\end{equation}
where $V$ is the potential on which $\vb*{\phi}$ slowly rolls due to the Hubble friction as long as the slow-roll parameter $\varepsilon \equiv \dd H^{-1} / \dd t$ is less than unity and thus $H^{2} \simeq V / 3 M_{\rm P}^{2}$ holds. The noise $\vb*{\xi}$, referred to as \textit{stochastic kick} in literature, comes from small-scale modes and randomly shifts the way the large-scale fields evolve. The Bunch--Davies vacuum initial condition implies that $\expval{ \vb*{\xi} (N) } = \vb*{0}$ with Gaussian statistics, and it is normalised in such a way that $\expval{ \vb*{\xi} (N_{1}) \otimes \vb*{\xi} (N_{2}) } = \delta_{\rm D} (N_{1} - N_{2}) \mathbbm{1}_{d \times d}$. 

With the nondimensionalised fields $\vb*{x} \equiv \vb*{\phi} / M_{\rm P}$, the rescaled potential is defined by $v (\vb*{x}) \equiv V (\vb*{\phi}) / 12 \pi^{2} M_{\rm P}^{4}$, and a derivative that acts on $v$ should be understood as $\nabla v = \partial v / \partial \vb*{x}$, to be used hereafter. Corresponding to the stochastic dynamics \eqref{eq:lan}, the distribution $f (\vb*{x}, \, N)$ of $\vb*{x}$ at a given time $N$ obeys the Fokker--Planck equation, 
\begin{equation}
	\pdv{f}{N} 
	= \qty[ 
        \nabla \cdot \frac{ \nabla v (\vb*{x}) }{v (\vb*{x})} 
        + \nabla^{2} \frac{v (\vb*{x})}{2} 
    ] f  
	\equiv \mathcal{L}_{\rm FP} f 
	\,\, . 
	\label{eq:fp}
\end{equation}
To obtain its solution, it must be supplemented by two boundary conditions besides an initial condition. Given that inflation terminates when the slow-roll condition is violated, an absorbing boundary is introduced by $C \equiv \qty{ \vb*{x} \mid \varepsilon (\vb*{x}) = 1 }$. A reflective boundary is also located at the other side in order to keep the fields below the Planck energy scale. 

Every stochastic trajectory generated by Eq.~(\ref{eq:lan}) realises a different duration of inflation. This promotes the number of $e$-folds $\mathcal{N}$, measured from an initial location $\vb*{x}$ until it hits $C$, to a stochastic number called the first-passage time. The distribution of $\mathcal{N}$ denoted $f_{\rm FPT} (\vb*{x}, \, \mathcal{N})$ satisfies $f_{\rm FPT} (\vb*{x} \in C, \, \mathcal{N}) = \delta_{\rm D} (\mathcal{N})$ as no finite time can be elapsed if $\vb*{x}$ is on the terminating surface from the beginning, and $f_{\rm FPT} (\vb*{x} \notin C, \, \mathcal{N} = 0) = 0$ since a finite number of $e$-folds is necessarily elapsed if the initial and final surfaces are not identical. It follows the adjoint Fokker--Planck equation, 
\begin{equation}
    \pdv{f_{\rm FPT}}{\mathcal{N}} 
    = \qty[ 
        - \frac{\nabla v (\vb*{x})}{v (\vb*{x})} \cdot \nabla 
        + \frac{v (\vb*{x})}{2} \nabla^{2} 
    ] f_{\rm FPT} 
    = \mathcal{L}_{\rm FP}^{\dagger} f_{\rm FPT} 
    \,\, . 
    \label{eq:fp_adj}
\end{equation}
This gives rise to the equation for the statistical moments of $\mathcal{N}$, 
\begin{equation}
	\qty[ \frac{v (\vb*{x})}{2} \nabla^{2} - \frac{ \nabla v (\vb*{x}) }{ v (\vb*{x}) } \cdot \nabla ] 
	\expval{ \mathcal{N}^{n} } (\vb*{x}) 
	= - n \expval{ \mathcal{N}^{n-1} } (\vb*{x}) 
	\,\, . 
	\label{eq:statmom_rec}
\end{equation}
Note that $\vb*{x}$ in Eq.~(\ref{eq:statmom_rec}) is the \textit{initial} location, for which each statistical moment is recursively determined from $\expval{ \mathcal{N}^{0} } = 1$. 

While it is analytically and even numerically challenging to solve those equations for a given potential, a class of $O (d)$-symmetric potentials (see \textit{e.g.}~\cite{Liddle:1998jc, Byrnes:2005th}) is of particular interest, for which the formal but analytical solution to Eq.~(\ref{eq:statmom_rec})~\cite{Assadullahi:2016gkk, Vennin:2016wnk} can be derived. 
For $n = 1$, one obtains 
\begin{equation}
	\expval{ \mathcal{N} } (r) 
	= \int_{r_{-}}^{r} \frac{\dd x}{x^{d-1}} \, e^{- 2 / v(x)} 
	\int_{x}^{r_{+}} \dd y \, y^{d-1} \frac{2}{v (y)} e^{2 / v (y)}
	\,\, . 
	\label{eq:mean_dint}
\end{equation}
Here, $r \equiv \norm{ \vb*{x} }$ only on which an $O (d)$-symmetric potential $v (\vb*{x}) = v (r)$ depends and, $r_{-}$ and $r_{+}$ are the absorbing and reflective boundaries respectively. In particular, $r_{-} = p / \sqrt{2}$ for the monomial potential, $v (r) = v_{0} r^{p}$. 

\vspace{8pt}
\noindent\textit{\bfseries Number of $e$-folds in small-noise regime.}---The regime of physical interest is when the stochastic effects are small since $v \ll 1$ in most scenarios. In such cases, Eq.~(\ref{eq:mean_dint}) can perturbatively be expanded in powers of $v$, resulting in 
\begin{equation}
	\expval{ \mathcal{N} } (r) 
	= \sum_{k = 0}^{\infty} \expval{ \mathcal{N} }^{(k)} (r) 
	\,\, , 
\end{equation}
where $\expval{ \mathcal{N} }^{(k)} = \mathcal{O} (v^{k})$. The integration over $y$ in Eq.~\eqref{eq:mean_dint} contains the two contributions of the dynamics of the field, the descending and ascending behaviours from $x \leq y \leq r$ and $r \leq y \leq r_{+}$ respectively. 
Meanwhile, the dominant contribution lies around $y = x$ in the small-noise regime since the field tends to descend unless a rare noise is realised. This motivates us to expand the integrand as $v (x) / v (y) = 1 - \qty[ v' (x) / v (x) ] (y - x) + \mathcal{O} ( (y - x)^{2} )$ as well as $e^{2 / v (y)}$, and at leading order one obtains~\cite{Assadullahi:2016gkk}
\begin{equation}
	\expval{ \mathcal{N} }^{(0)} (r) 
	= \int_{r_{-}}^{r} \dd x \, \frac{v (x)}{v' (x)} 
	\,\, . 
	\label{eq:meano0}
\end{equation}
The classical formula $\expval{ \mathcal{N} }^{(0)} = (r^{2} - r_{-}^{2}) / 2 p$ for the monomial potential $V \propto \norm{\vb*{\phi}}^{p}$ is indeed recovered. Since we consider a model with rotational symmetry, the field-space trajectory is along the straight line connecting the origin and the initial location $r$ in the absence of the stochastic noise. This prohibits the elapsed number of $e$-folds from depending on $d$. 
The explicit $d$-dependence is however observed from the next-to-leading order, 
\begin{equation}
	\expval{ \mathcal{N} }^{(1)} (r) 
	= \int_{r_{-}}^{r} \dd x \, v \qty( \frac{v}{v'} )^{2} \qty( 
		\frac{d-1}{2 x} + \frac{1}{2} \frac{v'}{v} - \frac{1}{2} \frac{v''}{v'} 
	) 
	\,\, . 
	\label{eq:meano1}
\end{equation}
When $d = 1$, Eq.~(\ref{eq:meano1}) reproduces the previously obtained single-field result~\cite{Vennin:2015hra}. For arbitrary $d > 1$, the $d$-dependent term contributes as it tends to increase the number of $e$-folds, which originates from the motion of the field deviating from the classical trajectory due to the stochastic noise. The higher-order terms can also be obtained, a detailed derivation of which is presented in our accompanying paper~\cite{inprep_tt} together with the variance and observables. For instance, the second-order correction reads 
\begin{widetext}
	\begin{equation}
		\expval{ \mathcal{N} }^{(2)} (r) 
		= \int_{r_{-}}^{r} \dd x \, v^{2} \qty( \frac{v}{v'} )^{3} \qty[ 
			\frac{(d-1) (d-2)}{4 x^{2}} + \frac{d-1}{x} \qty( \frac{v'}{v} - \frac{3}{4} \frac{v''}{v} ) 
			+ \frac{1}{2} \qty( \frac{v'}{v} )^{2} - \frac{v''}{v} 
			+ \frac{3}{4} \qty( \frac{v''}{v'} )^{2} - \frac{1}{4} \frac{v'''}{v'}  
		] 
		\,\, . 
		\label{eq:meano2}
	\end{equation}
\end{widetext}
It can be observed that there are two terms that depend on $d$ at the second order, one depends linearly and the other does quadratically. 
Neither of them does exist in single-field situations ($d = 1$) or if the stochastic effects are not taken into account. 

\begin{figure}
	\centering
	\includegraphics[width = 0.995\linewidth]{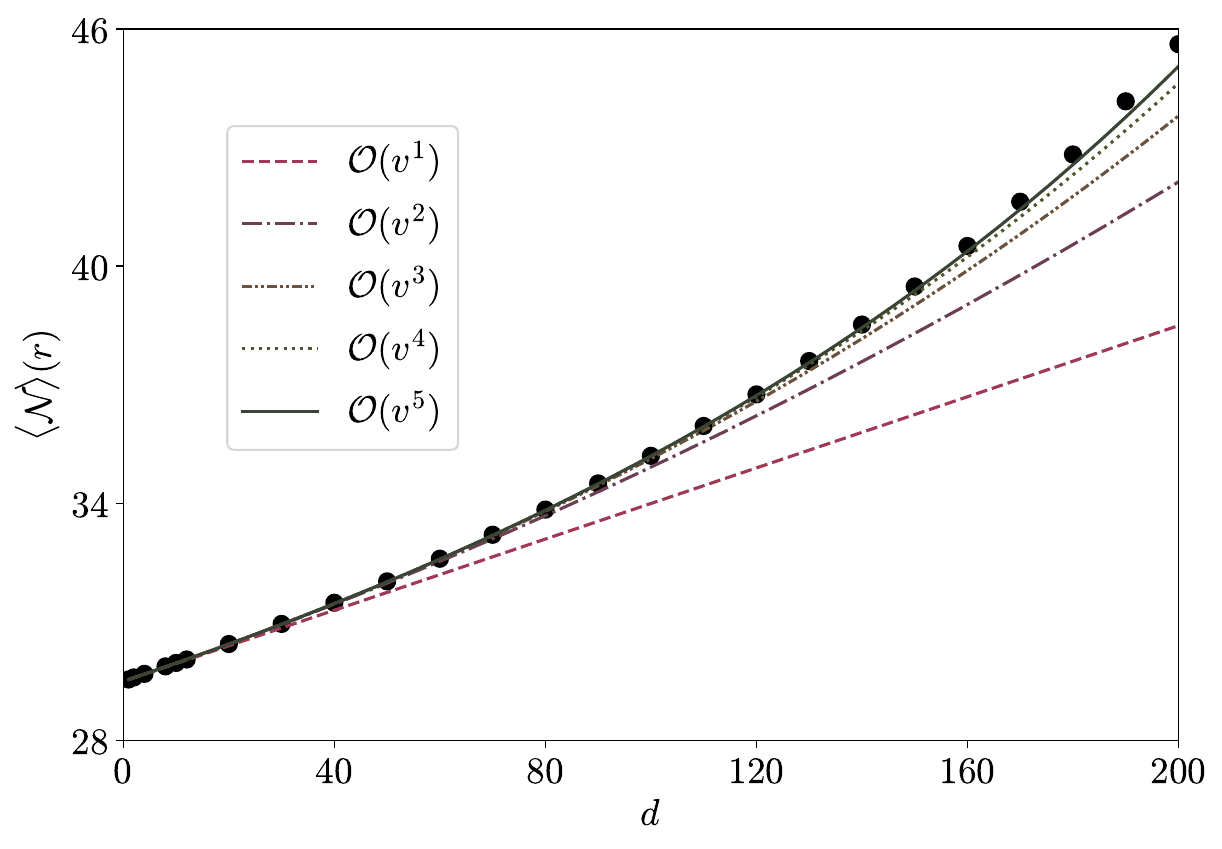}
	\caption{
		The mean number of $e$-folds 
        as a function of the number of fields $d$ for the monomial potential with $p=2$, 
        $v_{0} = 10^{-4}$, and $r = \sqrt{4 \times 30}$~\cite{supp_ref-1}. 
        The lines are the predictions from the formula~(\ref{eq:meanogen}) truncating at different orders. 
		Each dot represents the results from numerical simulations, in which $10^{5}$ stochastic realisations were generated by Eq.~(\ref{eq:r_lan}) and then averaged. 
	}
	\label{fig:numres}
\end{figure}

The expansion can in principle be performed in arbitrary order. While the general expression without specifying the potential becomes complicated as the order increases, it is simple for a class of the monomial potential $V\propto \norm{\vb*{\phi}}^p$, and reads 
\begin{equation}
	\expval{ \mathcal{N} } (r) 
	= \sum_{k = 0}^{\infty} \qty( \frac{v_0}{2} )^{k} \frac{ \Gamma (d/p + k) }{ p \Gamma (d/p) } \frac{r^{k p + 2} - r_{-}^{k p + 2}}{k p + 2} 
	\,\, . 
	\label{eq:meanogen}
\end{equation}
This is one of the main results of the present article. Figure~\ref{fig:numres} shows Eq.~(\ref{eq:meanogen}) up to fifth orders, with which the numerically obtained $e$-folds are well consistent if a sufficient order is kept for every $d$. The initial condition is imposed by $r = \sqrt{4 \times 30}$ for all the $d$'s simulated, \textit{i.e.}~the strength of the stochastic kick is common regardless of the number of fields. More and more terms are needed when $d$ gets larger though Eq.~(\ref{eq:meanogen}) is a perturbative expansion for small $v_{0}$. Also, each term contributes to $\expval{ \mathcal{N} }$ in the way that it increases the duration of inflation, which, at all-order level for arbitrary $d$, generalises and confirms the single-field result in~\cite{Vennin:2015hra}. 

\vspace{8pt}
\noindent\textit{\bfseries Condition for successful end of inflation.}---The formula~(\ref{eq:meanogen}) not only provides the analytical expression of $\expval{ \mathcal{N} }$ in small-noise regime, but also tells us more. Since it is an asymptotic series, there exists $k = \widetilde{k}$ below which the series well describes the actual value of $\expval{ \mathcal{N} }$. Now, for a fixed $k \leq \widetilde{k}$ and for $d \gg 1$, the $k$-th order term behaves as $k \expval{ \mathcal{N} }^{(k)} \approx \qty[ (v_{0}/2) \qty(d/p) r^{p} ]^{k} (r/p)^{2}$. A critical surface can therefore be found at $(v_{0}/2) \qty(d/p) r^{p} = 1$ that discriminates the convergence or divergence of the series. This implies that, in addition to $v_{0} \ll 1$ required in the small-noise regime, the number of fields $d$ and the initial location $r$ altogether determine the reliability of perturbative expansion~(\ref{eq:meanogen}). 

The existence of the convergence boundary can generically be understood as follows. In $O (d)$-symmetric models, the dynamics of the $d$ fields can be recast into the single equation for the radial displacement~\cite{Assadullahi:2016gkk, Tada:2023fvd}, 
\begin{equation}
	\dv{r}{N} 
	= - \frac{v' (r)}{v (r)} + \frac{v (r)}{2} \frac{d-1}{r} + \sqrt{ v (r) } \, \xi (N) 
	\,\, .  
	\label{eq:r_lan}
\end{equation}
In the classical limit, the last two terms are absent and the trajectory is given by $r^{2} - r_{-}^{2} = 2 p N$ for the monomial model. However, once the stochastic effects are provided, the noise-induced, centrifugal deterministic force affects the dynamics as well as the stochastic noise. In particular, the centrifugal-force term is peculiar to multi-field models, which originates from the motion that deviates from the classical trajectory due to the stochastic kicks. It vanishes when $d = 1$ and $\vb*{\phi}$ basically rolls down on the potential being exposed to the noise. When there are two or more fields, $\vb*{\phi}$ not only descend but can also ascend the potential even if $v$ is small. These two forces are balanced when the condition $v' (r) / v (r) = [ v (r) / 2 ] (d-1) / r$ holds, which is valid for generic $O(d)$-symmetric potentials, and gives the critical radius mentioned above.

\begin{figure}
	\centering
	\includegraphics[width = 0.995\linewidth]{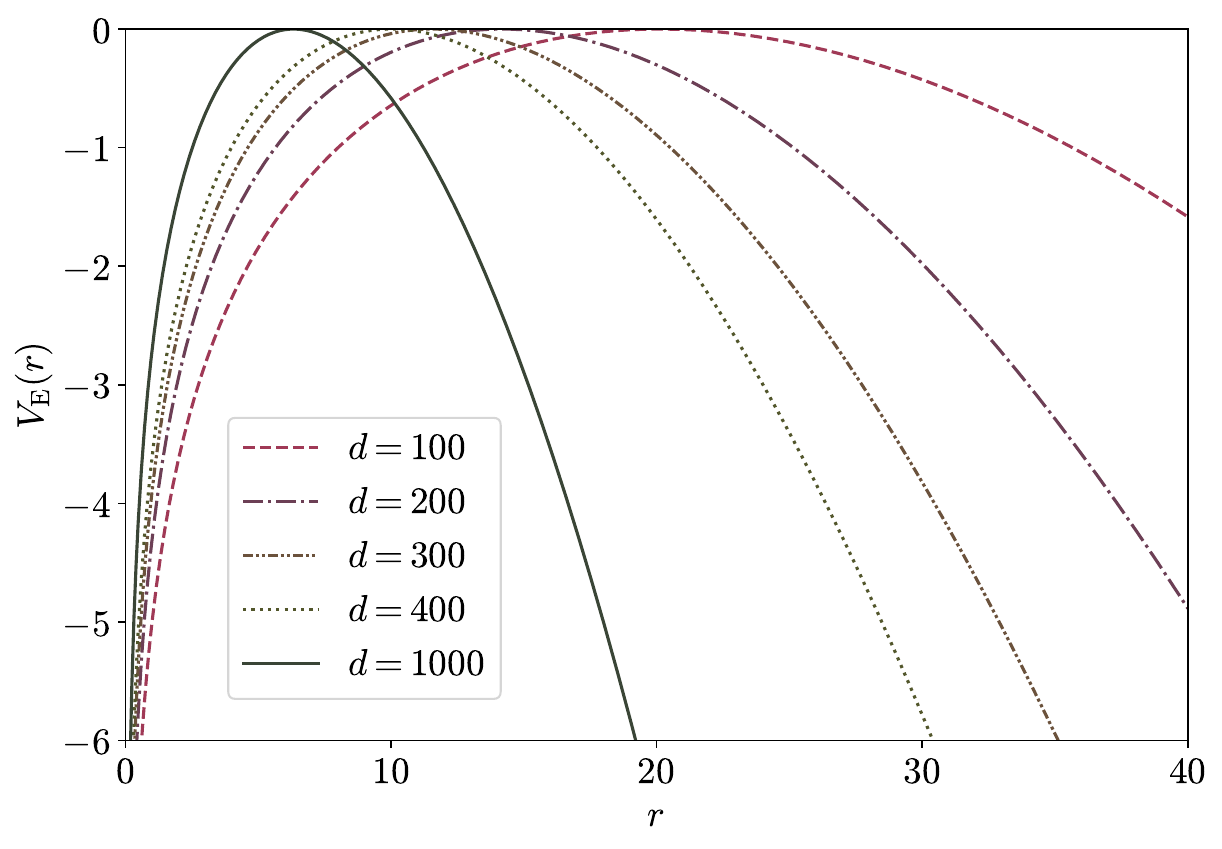}
	\caption{
		The effective potential defined in Eq.~(\ref{eq:efpot}) for $v (r) = v_{0} r^{2}$ and $v_{0} = 10^{-4}$. 
		The offset is fixed such that $V_{\rm E}$ vanishes at its maximum. 
	}
	\label{fig:veff}
\end{figure}

Let us introduce the effective potential defined by 
\begin{equation}
	V_{\rm E} (r)  
	\equiv \ln \qty[ v (r) ] - \int \dd r \,  \frac{v (r)}{2} \frac{d-1}{r} 
	\,\, . 
	\label{eq:efpot}
\end{equation}
Figure~\ref{fig:veff} shows Eq.~(\ref{eq:efpot}) for the quadratic monomial potential $v (r) = v_{0} r^{2}$ with $v_{0} = 10^{-4}$, and for several choices of the number of fields. This behaviour tells us the deterministic (not ``classical'') motion in the absence of the pure noise~\textit{i.e.}~$\xi$ in Eq.~(\ref{eq:r_lan}). It can be seen that not only $v_{0}$ and the initial location but the number of fields $d$ are all relevant, even under $O (d)$ symmetry. When $d = \mathcal{O} (1 - 10)$, there is a wide range for the initial displacement of the field that would give rise to a finite number of $e$-folds, namely, a successful end of inflation. However, when $d = \mathcal{O} (10^{3})$ or more fields are present, only a narrow range of $r$ allows us to realise the observable Universe, or otherwise $\expval{ \mathcal{N} }$ becomes infinite to result in eternal inflation. In particular, inflation inevitably becomes eternal in $d \to \infty$ limit. 

The criterion for inflation to terminate with a finite $\expval{ \mathcal{N} }$ is given by $r \lesssim \widetilde{r} \equiv \qty( 2 p / v_{0} d)^{1/p}$ for the initial location. This can equivalently be expressed as 
\begin{equation}
    d \lesssim \widetilde{d} \equiv \frac{2 p}{v_{0} r^{p}} 
    \,\, , 
\end{equation}
which puts an upper bound for the number of fields to realise our observable Universe for a given initial location. It should be noted that, even in those cases, a random noise sometimes drives the field in the $r > \widetilde{r}$ region to continuously ascend the potential, and hence inflation never ends. For instance, the upper bound is evaluated as $\widetilde{d} \approx 330$ for the case with $p=2$, $v_0 =10^{-4}$, and $r = \sqrt{4 \times 30}$, which corresponds to the values assumed in Fig.~\ref{fig:numres}. In the figure, all the dots are such that no realisation fails to end inflation since $d \ll \widetilde{d}$ and hence the classical drift dominates over the other two forces. On the other hand, if $d$ is smaller than but close to $\widetilde{d}$, the fields $\vb*{\phi}$ can, due to the stochastic noise, easily explore the region where inflation unavoidably becomes infinite. 

The reduced equation of motion (\ref{eq:r_lan}) also enables us to study the number of $e$-folds deterministically elapsed, in the regime where the noise dominates over neither the classical drift nor the diffusion-induced centrifugal force. In such cases, Eq.~(\ref{eq:r_lan}) can analytically be solved for $p = 2$ to give $\qty[ v_{0} (d-1) r^{2} / 4 ] \qty[ 1 - (\sqrt{2} / r)^{2} e^{- v_{0} (d-1) N} ] = 1 - e^{- v_{0} (d-1) N}$. This gives the trajectory of $r$, both inside and outside the convergence boundary. The deterministic part of the number of $e$-folds then reads 
\begin{equation}
	N_{d} (r) 
	= \frac{1}{v_{0} (d-1)} \ln \left[ 
		\frac{ 
			1 - \displaystyle \frac{v_{0} r^{2} (d-1)}{4} \biggl( \frac{\sqrt{2}}{r} \biggr)^{2} 
		}{ 
			1 - \displaystyle \frac{v_{0} r^{2} (d-1)}{4} 
		} 
	\right] \,\, . 
\end{equation}
This formula is valid as long as $r < \widetilde{r}$, and is along the dots in Fig.~\ref{fig:numres}. However, it diverges as $d \to \widetilde{d}$ as can be confirmed from the fact that the denominator in the logarithm approaches zero, which is consistent with the critical radius outside which the perturbative expansion~(\ref{eq:meanogen}) breaks down. 

\begin{figure}[!h]
	\centering
	\includegraphics[width = 0.995\linewidth]{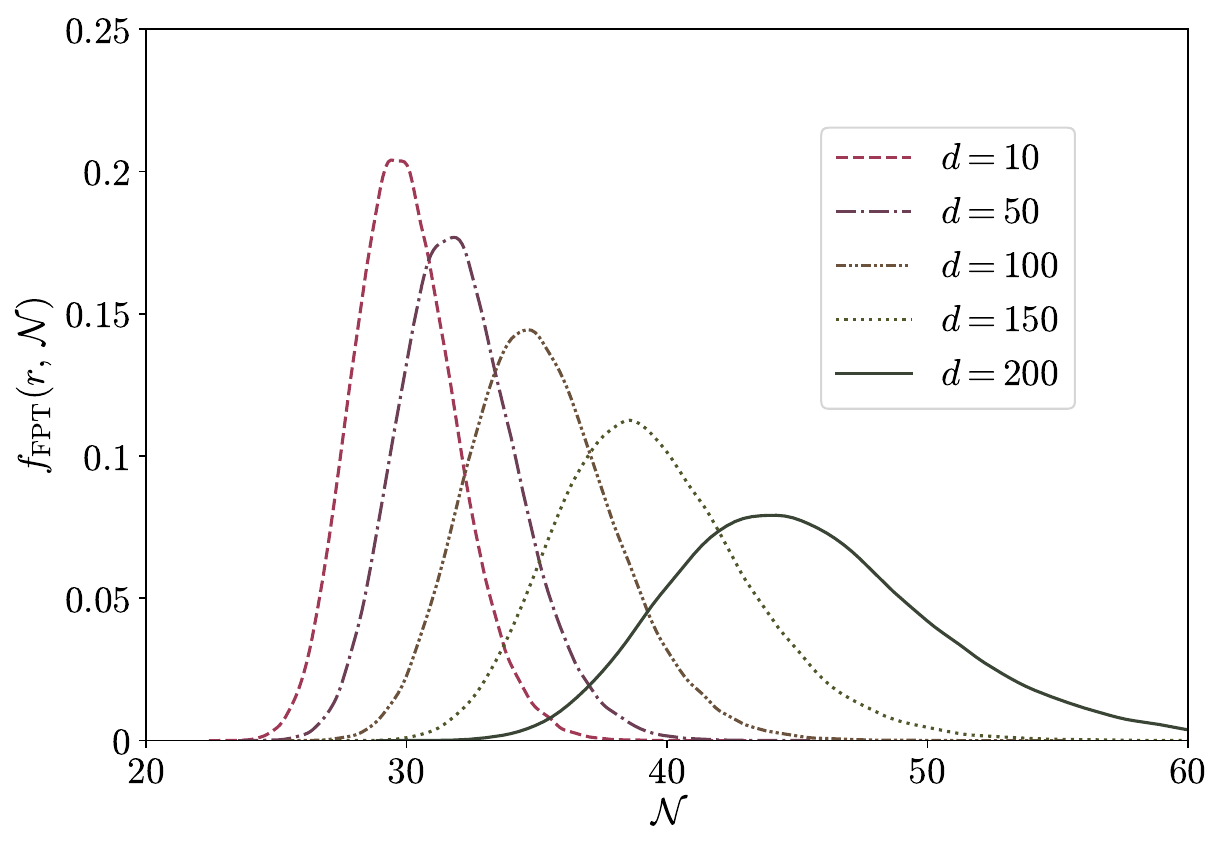}
	\caption{
		The reconstructed first-passage-time distribution for several $d$'s, which gives the solution to Eq.~(\ref{eq:fp_adj}). 
    	The parameters implemented are the same as those in Fig.~\ref{fig:numres}, see also~\cite{supp_ref-1}. 
	}
	\label{fig:numfpt}
\end{figure}

\vspace{8pt}
\noindent\textit{\bfseries Stochastic effects in variance of $e$-folds.}---The stochastic effects on the mean number of $e$-folds are focussed on until now, but a large number of stochastic realisations numerically generated to confirm the formula (\ref{eq:meanogen}) (see also Fig.~\ref{fig:numres}) also enables to reconstruct the first-passage-time distribution and higher-order statistical moments, the former of which is displayed in Fig.~\ref{fig:numfpt}. 

It can be observed that, not only the mean number of $e$-folds becomes larger, but also the variance around its mean value, $\delta \mathcal{N}^{2} (r) \equiv \expval{ \mathcal{N}^{2} } (r) - \expval{ \mathcal{N} }^{2} (r)$, gets larger, as more and more fields are prevalent. In addition to this, one can notice that the distribution deviates from Gaussian providing a heavier tail. Though similar behaviours can occur when the stochastic noise is large in single-field cases~\cite{Pattison:2017mbe, Ezquiaga:2019ftu, Pattison:2021oen}, the present outcomes are solely due to the noise-induced force peculiar to multi-field models. It can substantially affect the dynamics when a large number of fields drives inflation to realise more stochastic excursion, even when the noise itself is small. Quantitative discussion including the parametric dependence of the variance is presented in our upcoming paper~\cite{inprep_tt}. 

\vspace{8pt}
\noindent\textit{\bfseries Conclusion.}---The stochastic effects on the mean number of $e$-folds in multi-field setups were investigated. In small-noise regimes, the perturbative expansion of $\expval{ \mathcal{N} }$ at arbitrary order was presented. The extension of the mean duration of inflation was both perturbatively and numerically demonstrated for the monomial potential, yet this outcome would be found in a generic model without $O (d)$ symmetry where $\expval{ \mathcal{N} }$ is also affected by the stochastic kick in angular directions. In such cases, quantitative studies identifying the dependence of the critical surface on \textit{e.g.}~mass hierarchies call for fully numerical simulations, and are of great importance to be revealed in future work. 

The perturbative calculations of $\expval{ \mathcal{N} }$ unveiled the existence of a maximum number of fields, which depends on the setup of the model such as the initial field values and the potential itself. It originates from the balance condition between the two deterministic forces, and its existence is therefore generic. This means that the mean duration of inflation remains finite if only a few fields are there, whereas inflation never ends when a large number of fields are prevailing. In conclusion, the number of fields matters as to whether the theory can describe our Universe: More fields are indeed different. 

\vspace{8pt}
\noindent\textit{\bfseries Acknowledgements.}---This work was supported by JSPS KAKENHI Grant Numbers 23K17691 (TT) and 24K22877 (KT), and MEXT KAKENHI Grant Number 23H04515 (TT). 

\normalem
\bibliography{apssamp}

\begin{thebibliography}{34}
\bibitem [{sup()}]{supp_ref-1}%
  \BibitemOpen
\bibfield  {title} {  }\href@noop {} {}\bibinfo {note} {Our qualitative
  arguments do not depend on the specific choice of those
  parameters.}\BibitemShut {Stop}%
\end{thebibliography}

\end{document}